\documentstyle[12pt,aaspp4]{article}
\newcommand{\postscript}[2]
 {\setlength{\epsfxsize}{#2\hsize}
   \centerline{\epsfbox{#1}}}

\begin{document}

\title{Detection of Stellar Spots from the Observations of 
       Caustic-Crossing Binary-Lens Gravitational Microlensing Events}

\bigskip
\bigskip
\author{Cheongho Han}
\smallskip
\affil{Department of Astronomy \& Space Science, \\
       Chungbuk National University, Chongju, Korea 361-763 \\
       cheongho@astronomy.chungbuk.ac.kr}
\authoremail{cheongho@ast.chungbuk.ac.kr}

\author{Ho-Il Kim}
\smallskip
\affil{Korea Astronomy Observatory, \\
       Taejon, Korea 305-348 \\
       hikim@hanul.issa.re.kr}
\authoremail{hikim@hanul.issa.re.kr}

\author{Kyongae Chang}
\smallskip
\affil{Department of Physics, \\
       Chongju University, Chongju, Korea 360-764 \\
       kchang@alpha94.chongju.ac.kr}
\authoremail{kchang@alpha94.chongju.ac.kr}

\author{Sung-Hong Park}
\smallskip
\affil{Department of Astronomy \& Space Science, \\
       Chungbuk National University, Chongju, Korea 361-763 \\
       parksh@astronomy.chungbuk.ac.kr}
\authoremail{parksh@ast.chungbuk.ac.kr}

\bigskip

\begin{abstract}
Recently, Heyrovsk\'y \& Sasselov (1999) investigated the sensitivity of 
{\it single-lens} gravitational microlensing event light curves to small 
spots and found that during source transit events spots can cause deviations 
in amplification larger than 2\%, and thus be detectable.  In this paper, 
we explore the feasibility of spot detection from the observations of 
{\it caustic-crossing binary-lens} microlensing events instead of single-lens 
events.  For this we investigate the sensitivity of binary-lens event light 
curves to spots and compare it to that of single-lens events. From this 
investigation, we find that during caustic crossings the fractional 
amplification deviations of microlensing light curves from those of spotless 
source events are equivalent to the deviations of single-lens events, 
implying that spots can also be detected with a similar photometric precision 
to that required for spot detection by observing single-lens events.  We 
discuss the relative advantages of observing caustic-crossing binary-lens 
events over the observations of single-lens events in detecting stellar spots.
\end{abstract}


\centerline{submitted to {\it Monthly Notices of the Royal Astronomical
Society}: Jul 19, 1999}
\centerline{Preprint: CNU-A\&SS-07/99}
\clearpage

\section{Introduction}

Massive searches for gravitational microlensing events by monitoring transient 
brightening of source stars located in the Galactic bulge and the Magellanic 
Clouds have been and are being carried out by several groups (EROS: Aubourg 
et al.\ 1993; MACHO: Alcock et al.\ 1993; OGLE: Udalski et al.\ 1993; DUO: 
Alard \& Guibert 1997).  These surveys have detected $\sim 400$ events to 
date (Stubbs 1999).

The light curve of a single-lens microlensing event (denoted by the subscript 
`s') with a point source (denoted by the subscript `0') is given by
$$
A_{{\rm s},0}= {u^2+2\over u(u^2+4)^{1/2}},
\eqno(1.1)
$$
where $u$ is the lens-source separation in units of the angular Einstein ring 
radius $\theta_{\rm E}$.  The angular Einstein ring radius is related to the 
physical parameters of the lens by
$$
\theta_{\rm E}=\left( { 4GM\over c^2}
{ D_{ls}\over D_{ol}D_{os}}\right)^{1/2},
\eqno(1.2)
$$
where $M$ is the lens mass and $D_{ol}$, $D_{ls}$, and $D_{os}$ are the 
separations between the observer, lens, and source star, respectively.  Typical 
main-sequence stars in the Galactic bulge have radii that subtend only 
$\lesssim 1$ $\mu$-arcsecond, while the angular Einstein ring radius of an 
event caused by a solar mass lens with $D_{ol}\sim 5$ kpc is $\theta_{\rm E}
\sim 0.3$ milli-arcsecond.  Therefore, equation (1.1) is a good approximation 
for majority of Galactic microlensing events.

However, for a very close lens-source impact event with a considerable source 
star radius such as subgiants and giants, the source can no longer be 
approximated by a point source.  For this case, different parts of the source 
are amplified by different amounts (differential amplification) due to the 
finite size of the source star and the resulting light curve deviates from the 
point-source one (Schneider \& Weiss 1986; Witt \& Mao 1994).  The light curve 
of an extended source event caused by a single lens is given by the 
intensity-weighted amplification averaged over the surface of the source star, 
i.e.\ 
$$
A_{\rm s}(r_\ast)={\int_0^{2\pi}\int_0^{r_\ast} I(r,\vartheta)
A_{{\rm s},0}(\left\vert {\bf r}-{\bf r}_{L}\right\vert) rdrd\vartheta 
\over \int_0^{2\pi}\int_0^{r_\ast} I(r,\vartheta) rdrd\vartheta },
\eqno(1.3)
$$
where $r_\ast$ is the radius of the source star, $I(r,\vartheta)$ is the 
surface intensity distribution of the source star, and the vectors 
${\bf r}_L$ and ${\bf r}=(r,\vartheta)$ represent the displacement vector 
of the center of the source star with respect to the lens and the orientation 
vector of a point on the source star surface with respect to the source 
center, respectively.  

By observing the distortions in microlensing light curves caused by the 
finite source effect, one can obtain useful information about both the lens 
and source star.  First, it was known that finite source effect can be used 
to determine $\theta_{\rm E}$, with which one can partially break the lens 
parameter degeneracy in the obtained Einstein time scale $t_{\rm E}$ (Gould 
1994; Nemiroff \& Wickramasinghe 1994; Maoz \& Gould 1994; Peng 1997).
Second, since different parts of the source star (with varying surface 
intensity and spectral energy distribution) are resolved at different times 
during an event, one can recover the intensity profile of the source (Witt 
1995; Loeb \& Sasselov 1995; Gould \& Welch 1996) and can probe stellar 
atmosphere (Valls-Gabaud 1994, 1998; Sasselov 1997; Gaudi \& Gould 1999) by 
taking a sequence of photometric and spectro-photometric measurements of the 
event.

Recently, Heyrovsk\'y \& Sasselov (1999) investigated the sensitivity of 
{\it single-lens} microlensing event light curves to small spots with radii 
$r_{\rm s}\lesssim 0.2$ of source radii.  From this investigation, they found 
that during source transit events spots can cause deviations in amplification
larger than 2\%, and thus be detectable.  In this paper, we explore the 
feasibility of spot detection from the observations of {\it caustic-crossing 
binary-lens} microlensing events instead of single-lens events.  For this we 
investigate the sensitivity of binary-lens event light curves to spots and 
compare it to that of single-lens events. From this investigation, we find 
that during caustic crossings the fractional amplification deviations of 
microlensing light curves from those of spotless source events are equivalent 
to the deviations of single-lens events, implying that spots can also be 
detected with a similar photometric precision to that required for spot 
detection by observing single-lens events.  We discuss the relative advantages 
of observing caustic-crossing binary-lens events over the observations of 
single-lens events in detecting stellar spots.

\section{Single-Lens Events for Finite Source Stars with Spots}

If the surface of a source star is maculated by a spot, the light curve of 
a single-lens microlensing event becomes
$$
A_{\rm s,spot}=
{
\int_0^{2\pi}\int_0^{r_\ast} I(r,\vartheta) A_{{\rm s},0} 
(\left\vert {\bf r}-{\bf r}_{L}\right\vert) rdrd\vartheta -
\int_{\Sigma_{\rm spot}} f({\bf r}') I({\bf r}') A_{{\rm s},0} 
(\left\vert {\bf r'}-{\bf r}_{L}\right\vert) d\Sigma_{\rm spot}
\over 
\int_0^{2\pi}\int_0^{r_\ast} I(r,\vartheta)\left[ 1-f(r,\vartheta)\right] 
rdrd\vartheta
},
\eqno(2.1)
$$
where ${\bf r}'$ is the orientation vector of a point on the surface of 
the spot with respect to the source center, $f({\bf r}')$ represents the 
fractional decrement in the surface intensity due to the spot, and the 
notation $\int_{\Sigma_{\rm spot}} \cdot\cdot\cdot\ d\Sigma_{\rm spot}$ 
represents the surface integral over the spot range of the source star.

Due to the presence of the spot, the event light curve deviates from that 
of a spotless event.  To see the pattern of the deviations in microlensing
light curves caused by spots and to explore the feasibility of spot detection 
by this method, we compute the fractional amplification deviation 
in the light curve from that of a spotless event, i.e.\  
$$
\epsilon_{\rm s}={\left\vert A_{\rm s,spot}-A_{\rm s}\right\vert
\over A_{\rm s}},
\eqno(2.2)
$$ 
by using equations (1.3) and (2.1).  For the computation of $\epsilon_{\rm s}$,
we assume a constant surface brightness of $I_\ast$ for the entire region of 
the source star outside the spot.  Partially, this is because the limb
darkening does not have significant effect on the light curve, but more 
importantly because we want to see the deviation caused sorely by the spot.  
The spot is modeled by a circular area with a radius $r_{\rm spot}$ and also 
has a uniform surface brightness $I_{\rm spot}$.  We test two cases of events 
for which the source stars have radii of $r_\ast=0.05 \theta_{\rm E}$ and 
$0.1 \theta_{\rm E}$.  For both cases, the spots have relative radii of 
$r_{\rm spot}/r_\ast=0.2$ with the surface brightness contrast parameter of 
${\cal C}=I_\ast/I_{\rm spot}=1/(1-f)=10$.  With these assumptions, the light 
curve in equation (2.1) is simplified into
$$
A_{\rm s,spot} (r_\ast, r_{\rm spot}, f)=
{
\int_0^{2\pi}\int_0^{r_\ast} A_{{\rm s},0} 
(\left\vert {\bf r}-{\bf r}_{L}\right\vert) rdrd\vartheta -
f \int_0^{2\pi}\int_0^{r_{\rm spot}} A_{{\rm s},0} 
(\left\vert {\bf r'}-{\bf r}_{L}\right\vert) r'dr'd\vartheta'
\over 
\pi (r_\ast^2-f r_{\rm spot}^2)
}.
\eqno(2.3)
$$

In the upper panels of Figure 1, we present the contours of the deviation 
$\epsilon_{\rm s}$ as a function of the {\it lens} position $(x_L,y_L)$.  In 
each panel, the two circles represent the source star (big empty circle 
centered at the origin) and the stellar spot on it (small filled circle), 
respectively.  The spot is located at $s=0.5 r_{\rm spot}$ from the center 
of the source star.  Contours are drawn with a spacing of 0.2\% from 
$\epsilon_{\rm s}=0.2\%$ and the regions with $\epsilon_{\rm s} \geq 1\%$ and 
$\epsilon_{\rm s} \geq 2\%$ are shaded by darkening gray tones.  In the lower 
panels, we also present several example light curves of events for source 
stars with (solid lines) and without spots (dotted lines) and the corresponding
lens trajectories (dot-long dashed lines) are marked in the upper panels.  
Each pair of the trajectory and the corresponding light curve are marked by 
the same number.  We note that our Figure 1 is equivalent to Figure 1 and 2 of 
Heyrovsk\'y \& Sasselov (1999), except that their adopted source size is 
$r_\ast=\theta_{\rm E}/13.23$ and all their lengths are scaled by the source 
size not by the angular Einstein ring radius.

From the figure, one finds that the deviation can be larger than 2\% as noted 
by Heyrovsk\'y \& Sasselov (1999), and thus spots can, in principle, be 
detectable.  However, the region for noticeable deviations (e.g.\ 
$\epsilon_{\rm s}\geq 2\%$) is confined to a very small localized area 
close to the spot.  Even for source-transit events, unless the lens almost 
directly crosses the spot, the deviations will not be big enough to notice 
the existence of the spot.  This implies that with a reasonable photometric 
precision, spots can be detected only for a very limited number of (almost) 
direct spot-transit events.

\section{Caustic-Crossing Binary-Lens Events for Finite Source Stars 
with Spots}

In previous section, we investigated the effect of stellar spots on the light 
curves of single-lens microlensing events.  In this section, we investigate 
how the spots of source stars affect the light curves of caustic-crossing 
binary-lens events.

When lengths are normalized to the {\it combined} Einstein ring radius 
$r_{\rm E}$, which is equivalent to the Einstein ring radius of a single lens 
with a mass equal to the total mass of the binary, the lens equation in complex
notations for a binary-lens system with a point source is given by
$$
\zeta = z + {m_{1} \over \bar{z}_{1}-\bar{z}} 
+ {m_{2} \over \bar{z}_{2}-\bar{z}},
\eqno(3.1)
$$
where $m_1$ and $m_2$ are the mass fractions of individual lenses (and thus 
$m_1+m_2=1$), $z_1$ and $z_2$ are the positions of the lenses, $\zeta = 
\xi +i\eta$ and $z=x+iy$ are the positions of the source and images, and 
$\bar{z}$ denotes the complex conjugate of $z$ (Witt 1990).  The amplification 
of each image, $A_i$, is given by the Jacobian of the transformation (3.1) 
evaluated at the images position, i.e.\  
$$
A_{{\rm b},0,i} = \left({1\over \vert {\rm det}\ J\vert} \right)_{z=z_i};
\qquad {\rm det}\ J = 1-{\partial\zeta\over\partial\bar{z}}
{\overline{\partial\zeta}\over\partial\bar{z}}.
\eqno(3.2)
$$
Then the total amplification of a binary-lens event (denoted by the subscript 
`b') with a point source is given by the sum of the amplifications of the 
individual images, i.e.\ $A_{{\rm b},0}=\sum_i A_{{\rm b},0,i}$.  The set of 
source positions with infinite amplifications, i.e.\ ${\rm det}\ J=0$, form 
closed curves called caustics.  Therefore, whenever a point source crosses the 
caustic, the amplification becomes formally infinity, producing a sharp peak 
in the light curve.  Since the caustics form a closed figure, the source 
transit occurs at least twice for a caustic-crossing binary-lens event.

The finite source effect also affects the light curves of binary-lens events.  
The light curve of a binary-lens event with a finite source, $A_{\rm b}$, is 
obtained in a similar fashion to the single-lens event case, i.e.\ the 
intensity-weighted amplification averaged over the surface of the source star 
as is in equation (1.3) except that the single-lens point-source amplification 
$A_{\rm s,0}$ should be replaced by the binary-lens amplification for a point 
source $A_{\rm b,0}$.  Due to the finite source effect, the observed 
amplification remains finite even during the caustic crossings.  

If a spot exists on the surface of a finite source, the light curve further
deviates from $A_{\rm b}$.  The amplification of the binary-lens event for 
a source star with a spot, $A_{\rm b,spot}$, is obtained by using equation 
(2.3), but with $A_{\rm b,0}$ instead of $A_{\rm s,0}$.  To see how the light 
curves of binary-lens events are affected by spots and to explore the 
feasibility of using binary-lens events for spot detection, we compute the 
fractional deviation in the amplification from that of a spotless event by
$$
\epsilon_{\rm b}= {\left\vert A_{\rm b,spot} - A_{\rm b}\right\vert
\over A_{\rm b} },
\eqno(3.3)
$$
and the result is presented in Figure 2.  In the upper panel of Figure 2, we 
present the contours of $\epsilon_{\rm b}$ in the vicinity of the caustics of 
an example binary-lens system with a binary separation (normalized by 
$\theta_{\rm E}$) and mass ratio of $a=1.0$ and $q=1.0$, respectively.  The 
closed figure (marked by thick solid curves) in each panel represents the 
caustics of the binary-lens system.  The contours are drawn at the levels of 
$\epsilon_{\rm b}=1\%$ and 2\% and the regions with $\epsilon_{\rm b}\geq 1\%$ 
and $\epsilon_{\rm b}\geq 2\%$ are shaded by darkening gray tones.  For direct 
comparison of the deviations to those of the single-lens events in Figure 1, 
we adopt the same radii of source stars and their spots, i.e.\ $r_\ast=0.05 
\theta_{\rm E}$ and $0.1\theta_{\rm E}$ and $r_{\rm spot} = 0.2r_\ast$, and 
the surface brightness contrast, i.e.\ ${\cal C}=10$.  Unlike the single-lens 
event cases, however, we place the spots at the center of the source stars, 
i.e.\ $s=0$, but we will discuss the dependency of the deviation 
$\epsilon_{\rm b}$ on the spot position in the following paragraph.  In the 
lower panels, we also present several light curves for source stars with 
(solid lines) and without spots (dotted lines).  The source star trajectories 
corresponding to individual light curves are represented by dot-long dashed 
lines in the upper panels and each pair of the light curve and trajectory 
are marked by the same number.

From the figure, one finds following patterns of $\epsilon_{\rm b}$.  First, 
significant deviations in amplification occur at near the regions along the 
caustics, implying that noticeable devaitions in light curves occur when the 
spot crosses the caustic.  If the spot is located at different positions on 
the source star, the regions for optimial spot detection will change because 
the spot will cross the caustic at a different time.  However, since the spot 
is confined within the small region of the source star, the change will be very 
slight.  Second, since the significant deviation region well surrounds most of 
the cautic lines, one can detect the deviation for nearly all caustic-crossing 
events regardless of the lens trajectories.  Third, while the center-to-limb 
surface intensity variation, which is another important stellar surface 
structure, produces very smooth deviations in the light curve (e.g.\ the 
Galactic bulge event MACHO 97-BLG-28, Albrow et al.\ 1999a), the deviations 
caused by the spot are bumpy.  Therefore, one can easily separate the 
deviations in the light curve caused by the spot.

\section{Single-Lens Versus Binary-Lens Events}

Although  both single-lens and binary-lens microlensing events can be used 
to detect stellar spots, observing caustic-crossing binary-lens events have
following relative advantages in detecting spots over the observations of 
source-transit single-lens events.

First, caustic-crossing binary-lens events are much more common than 
source-transit single-lens events.  Currently, total 11 candidate 
caustic-crossing binary-lens events have been reported.  These include MACHO 
LMC\#\hskip-1pt 1 (Dominik \& Hirshfeld 1994, 1996; Rhie \& Bennett 1996), 
OGLE\#\hskip-1pt 7 (Udalski et al.\ 1994), DUO\#\hskip-1pt 2 (Alard, Mao, \& 
Guibert 1995), 97-BLG-28 (Albrow et al.\ 1999a), 98-SMC-1 (Afonso et al.\ 1998;
Albrow et al.\ 1999b; Alcock et al.\ 1999), 96-BLG-3, 97-BLG-1, 97-BLG-41, 
98-BLG-12, 98-BLG-42, and 99-BLG-28, (http://darkstar.astro.washington.edu).  
On the other hand, only one candidate source-transit single-lens event has 
been reported (MACHO Alert 95-30, Alcock et al.\ 1997).  In addition, while 
one can detect the deviations caused by spots for nearly all caustic-crossing 
events, spot detection for a significant fraction of source-transit single-lens
events will be difficult because only almost direct spot-transit events will 
produce deviations large enough to detect spots.

Second, for a caustic-crossing binary-lens event, the deviations caused by 
a spot can be measured with precision and high time resolution from followup 
observations of the event.  The deviations in the light curve last only a few 
hours during the source star transit of the lens (for a single-lens event) or 
the caustic (for a binary-lens event).  Therefore, followup observations with 
high photometric precision and time resolution will be essential for the 
detection of stellar spots.  For a binary-lens event, the caustic crossing 
happens twice and thus although the first crossing was missed one can prepare
the  followup observations of the second crossing.  By contrast, since the 
source transit of a single-lens event cannot be predicted a priori as well as
never repeats, followup observations cannot be performed for an important 
fraction of these events.

\acknowledgements

\clearpage

\postscript{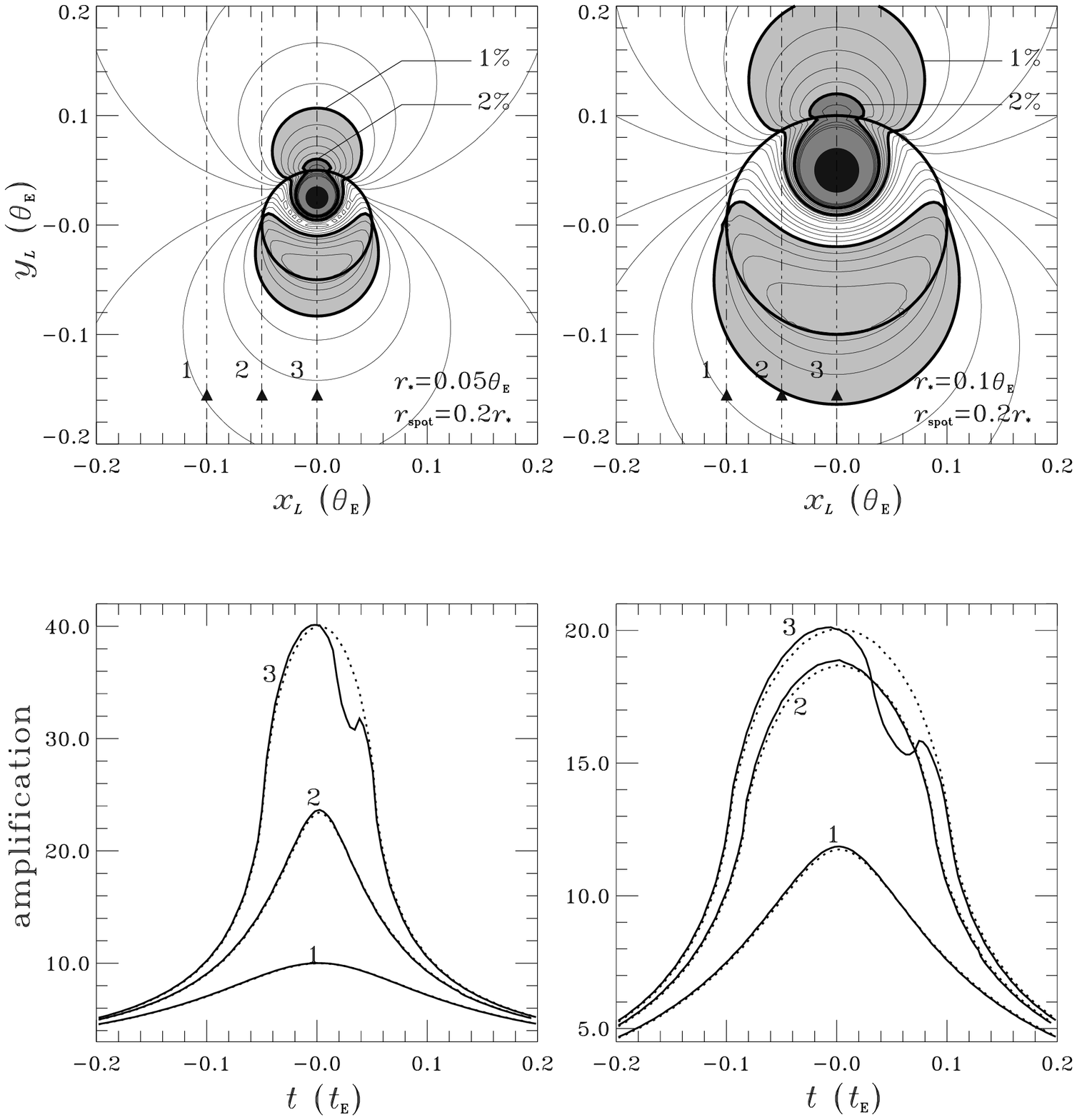}{0.87}
\vskip-5.0cm
\noindent
{\footnotesize {\bf Figure 1:}\ 
Upper panels: contours of the fractional amplification deviation 
$\epsilon_{\rm s}$ as a function of the {\it lens} position $(x_L,y_L)$ for 
single-lens microlensing events.  The two circles in each panel represent the 
source star (big empty circle centered at the origin) and the stellar spot on 
it (small filled circle), respectively.  The source stars have radii of 
$r_\ast=0.05\theta_{\rm E}$ and $0.1\theta_{\rm E}$.  For both cases, the 
spot has a relative radius of $r_{\rm s}/r_\ast =0.2$ and the adopted contrast 
parameter ${\cal C}= I_\ast/I_{\rm spot} =10$ for both cases.  Contours are 
drawn with a spacing of 0.2\% from $\epsilon_{\rm s}= 0.2\%$ and the regions 
with $\epsilon_{\rm s} \geq 1\%$ and $\epsilon_{\rm s} \geq 2\%$ are shaded 
by darkening gray tones.
Lower panels: the light curves (solid curves) of single-lens microlensing 
events for source stars with spots.  The lens trajectories corresponding to 
the individual light curves are represented by dot-long dashed lines in the 
upper panels and each pair of the light curve and trajectory are marked by 
the same number.  The dotted curves represent the light curves which are 
expected when the source stars have no spot.
}\clearpage

\postscript{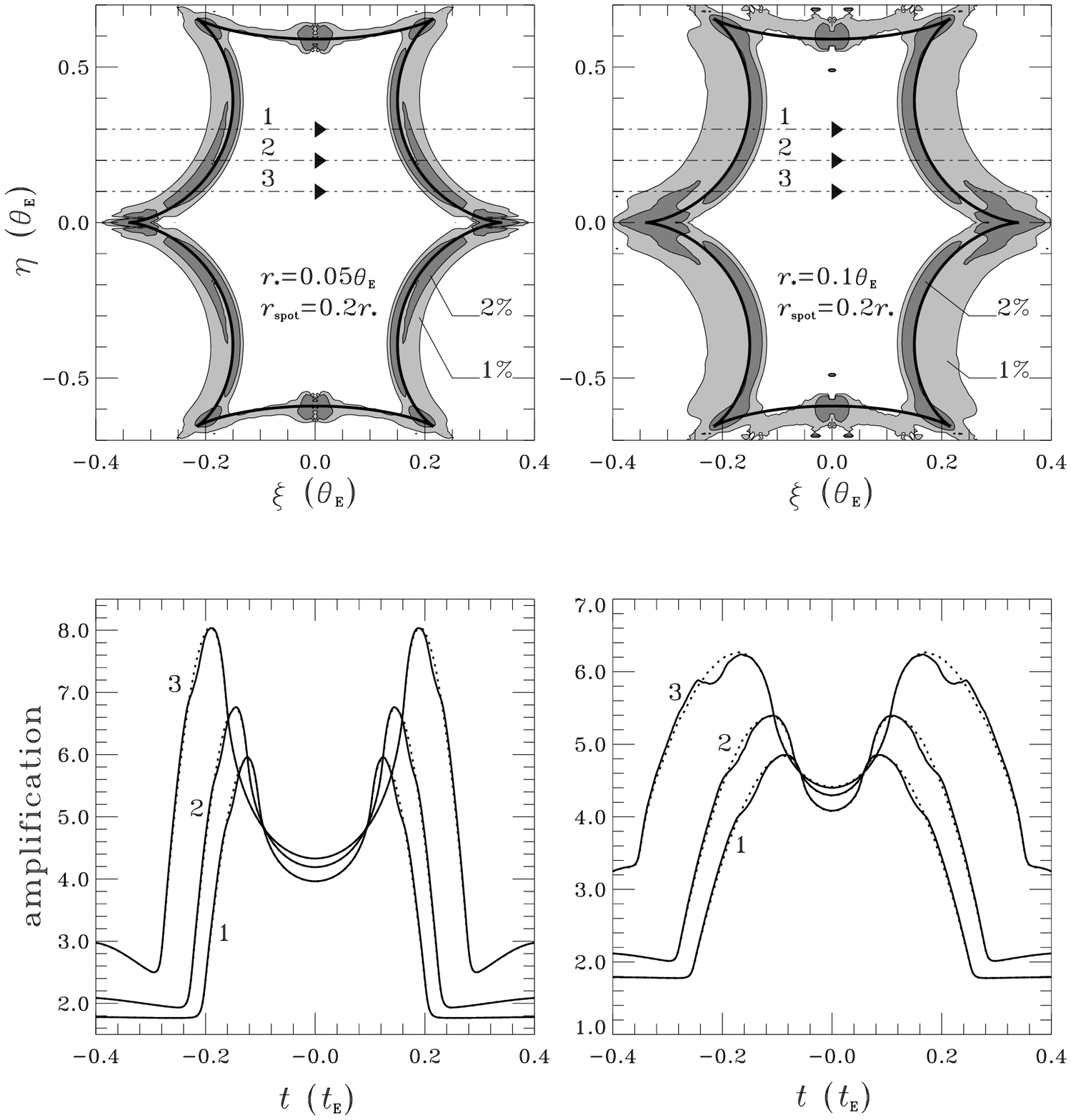}{0.87}
\vskip-5.0cm
\noindent
{\footnotesize {\bf Figure 2:}\ 
Upper panel:
contours of amplification deviation $\epsilon_{\rm b}$ as a function of the 
{\it source star} position $(\xi,\eta)$ for binary-lens microlensing events.  
The lens system is composed of equal mass lenses (i.e.\ mass ratio $q=1.0$) 
with a normalized binary separation of $a=1.0$.  The closed figure (marked 
by thick solid curves) in each panel represents the caustics of the binary-lens
system.  The contours are drawn at the levels of $\epsilon_{\rm b}=1\%$ and 
2\% and the regions with $\epsilon_{\rm b}\geq 1\%$ and 
$\epsilon_{\rm b}\geq 2\%$ are shaded by darkening gray tones.  The radii of 
source stars and their spots and the surface brightness contrast are same as 
in Figure 1.
Lower panel: the light curves (solid curves) of binary-lens microlensing 
events for source stars with spots.  The source star trajectories corresponding
to the individual light curves are represented by dot-long dashed lines in the 
upper panels and each pair of the light curve and trajectory are marked by the 
same number.  The dotted curves represent the light curves without spots.
}\clearpage

\end{document}